# Field Output correction factors of small static field for IBA Razor NanoChamber

D. Mateus[1,2,3], C. Greco[3], L. Peralta[1,4]

[1]Faculdade de Ciências da Universidade de Lisboa, Lisboa, Portugal
[2]Mercurius Health S.A, Lisboa, Portugal
[3]Fundação Champalimaud, Lisboa, Portugal,
[4]Laboratório de Instrumentação e Física Experimental de Partículas, Lisboa, Portugal

**Purpose:** The goal of this work is to present results of field output factors (OF) using an IBA CC003 (Razor NanoChamber) and compared these results with PTW 60019 (MicroDiamond) and IBA Razor Diode. The experimental results for IBA CC003 were also compared with Monte Carlo (MC) Simulation, using Penelope and Ulysses programs. In addition, field output correction factors ($k_{Q_{clin},Q_{msr}}^{f_{clin},f_{msr}}$) for IBA CC003 were derived with three different methods: 1) using PTW 60019 and IBA Razor as reference detectors; 2) comparison between MC and experimental measurements; and 3) using only MC.

**Material and Methods:** The beam collimation included in this study were 1) square field size between 10x10 and 0.5x0.5 cm$^2$ defined by the MLC and jaws and 2) cones of different diameters. For IBA CC003 it was determined the polarity and ion collection efficiency correction factors in parallel and perpendicular orientation.

**Results:** The results indicate 1) the variation of polarity effect with the field size is relevant for the determination of OF using IBA CC003, especially for parallel orientation; 2) there is no significant variation of the ion collection efficiency with the field size using IBA CC003 in parallel orientation; 3) OF differences between IBA CC003 and PTW 60019/IBA Razor and experimental and MC results increase with decreasing field size;

The $k_{Q_{clin},Q_{msr}}^{f_{clin},f_{msr}}$ results indicate 1) using the first and second method, $k_{Q_{clin},Q_{msr}}^{f_{clin},f_{msr}}$ increase with decreasing field size, which can be related with the influence of the volume effect and 2) using the third method, $k_{Q_{clin},Q_{msr}}^{f_{clin},f_{msr}}$ decrease with decreasing field size, which can be explained by the perturbation effect.

**Conclusions:** Our results demonstrate the need of applying $k_{Q_{clin},Q_{msr}}^{f_{clin},f_{msr}}$ for IBA CC003 for $S_{clin} \leq 1$ cm, to compensate for volume averaging and perturbations effects.

# 1. Introduction:

In the last years, the use of small fields (less than 2 cm of diameter) in Radiotherapy has increased with the use of advanced treatment techniques, such as, stereotactic radiosurgery (SRS), stereotactic body radiotherapy (SBRT), intensity modulated radiotherapy (IMRT) and volumetric modulated radiotherapy (VMAT). The diagnosis of small lesions has also contributed to this need, due to the use of image techniques, such as, magnetic resonance imaging (MRI) and positron emission tomography (PET).

The measurement of field output factors (OF) for MV small photon fields are subjected to large uncertainties, due to the challenges of small field dosimetry, because of the lack of electronic equilibrium, source occlusion and volume effect, and also the requirements of the detectors used in the measurements (IAEA TRS-483 2017, Aspradakis M et al. 2010, Das I et al. 2008). Thus, the small field dosimetry cannot be done with the conventional dosimeters and protocols (Andreo P et al. 2004).

Alfonso et al (2008) [5] proposed a new formalism for small and non-standard field dosimetry, introducing a new concept of using correction factor $k_{Q_{clin},Q_{msr}}^{f_{clin},f_{msr}}$ which correlates the differences between the clinical field size $f_{clin}$ and the machine-specific reference field size $f_{msr}$. The International Atomic Energy Agency (IAEA) in conjunction with American Association of Physicists in Medicine (AAPM) compiled the studies about small field dosimetry and developed the new formalism introduced by Alfonso and presented the TRS-483 Code of Practice (IAEA TRS-483 2017), which provides correction factors for different detectors and small field sizes.

In the market there are different kinds of detectors to perform measurements of small fields, which present advantages and disadvantages for the measurements. However, the ionization chambers are known as the workhorse of reference dosimetry, thus it would be ideal performing the small field dosimetry using these kinds of detectors.

The purpose of this study is to contribute with measurements for small field sizes with IBA Razor NanoChamber (IBA CC003) (Reggiori G et al. 2017, Razor NanoChamber 2017) using experimental data and Monte Carlo simulation for 6 MV and 10 MV with and without flattening filter beams. In addition, we seek to derive field output correction factors $k_{Q_{clin},Q_{msr}}^{f_{clin},f_{msr}}$ for IBA CC003 using three methods: 1) reference detectors (PTW 60019 and IBA Razor) with known output correction factors; 2) comparison between Monte Carlo simulation and experimental measurements; and 3) using only Monte Carlo to derive field output factors.

# 2. Material & Methods:

**2.1 Accelerators and photon beams:**

The dosimetric measurements were performed in three Varian machines (Varian Medical Systems, Palo Alto, CA, USA): two TrueBeam (TrueBeam1 and TrueBeam2) and one Edge. The TrueBeam1 and Edge have a high definition MLC (central leafs with 0.25 cm of thickness) and TrueBeam2 has Millennium 120 MLC (central leafs with 0.5 cm of thickness).

The nominal photon energies used were: 6 MV (with (6X) and without (6X-FFF) flattening filter) and 10 MV (with (10X) and without (10X-FFF) flattening filter). The dose-rate (MU/min) used was 600 MU/min and 800 MU/min for flattened and unflattened beams, respectively.

The measurement geometry consisted of an isocentric setup with source-to-surface distance (SDD) of 90 cm and a depth of 10 cm with gantry at 0°.

The beam collimation included in this study were 1) cones of different diameters (1.75, 1.5, 1.25, 1.0, 0.75 cm) with jaws opening of 5x5 cm$^2$; and 2) square field size of 10x10, 4x4, 3x3, 2x2, 1x1 and 0.5x0.5 cm$^2$ defined by the MLC with jaws positioning of 11x11, 5x5 and 4x4 cm$^2$ for square field size of 10x10, 4x4 and 3x3 cm$^2$, respectively, and 3x3 cm$^2$ for the rest square field sizes; and 3) square field size of 10x10, 4x4, 3x3, 2x2, 1.5x1.5, 1x1, 0.8x0.8 and 0.5x0.5 cm$^2$ defined by the jaws. The 10x10 cm$^2$ field size was used as reference field size for the OF determination.

## 2.2 Detectors:

In this work, the detectors used were the ionization chamber IBA CC003, the microDiamond detector PTW 60019 and the diode IBA Razor. The IBA CC003 (Reggiori G et al. 2017, Razor NanoChamber 2017) is a microchamber (active volume of 3.0 mm³), whose outer and inner electrodes are made of Shonka (C-552) plastic and graphite-EDM3, respectively. The outer electrode of Shonka, which corresponds to air equivalent plastic, has a diameter of 2.0 mm and a density of $\rho = 1.76\ g/cm^3$. The inner electrode of graphite has a diameter of 1.0 mm and a density of $\rho = 1.81\ g/cm^3$. The PTW60019 detector (PTW User Manual 2014), with an active volume of 0.004 mm³, consists of a single crystal intrinsic layer with thickness of 1 μm and a diameter of 2.2 mm. The IBA Razor diode (Razor Detector 2014) has an active volume of 0.6 mm in diameter and 20 μm of thickness. It is made with a n-type implant in p-type silicon and operates in photovoltaic mode.

The IBA CC003 was placed in two setups, with the chamber stem parallel and perpendicular to the beam axis and operated with a bias voltage of 300 V. The polarity effect for IBA CC003 was evaluated for the 4 energies used and its dependence on the field size was investigated for the two orientations of the ionization chamber. The polarity correction factor was measured, for each field size, according to Eq. (1):

$$k_{pol} = \frac{|M_{+300}| + |M_{-300}|}{2M_{+300}} \quad (1)$$

The ion collection efficiency was also investigated for the field sizes presented in this study. It was reported by different authors (Weinhous M S and Meli J A 1984, Hyun M A 2017, Zankowski C and Podgorsak E B 1998, Agostinelli S et al. 2008) that for very small volume ionization chambers, like IBA CC003, and high voltages the traditional two-voltage method is no longer adequate for definition of recombination effect ($k_s$). Instead of that should be applied a method proposed by Agostinelli (Agostinelli S et al. 2008), which evaluates the variation of $\frac{1}{Q}$ vs. $\frac{1}{V}$ (saturation curves) where $Q$ is the collected charge and $V$ is the applied voltage. Jaffé plots (Zankowski C and Podgorsak E B 1998) were created from the saturation curves. The saturation value of the collected charge $Q_{sat}$ was calculated as the inverse of the interception for $\frac{1}{V} \to 0$ of the linear regression curve. The linear fit was performed considering the linear part of the plot only (i.e. low voltages: 25, 50, 75, 100 $V$). Following this method, the ion collection efficiency correction factor was introduced as:

$$J_s = \frac{Q_{sat}}{Q_{300}} \quad (2)$$

where $Q_{300}$ is the collected charge at the normal operating voltage of 300 V. The variable $J_s$ corrects for both recombination and excess charges and reduces to the conventional recombination factor if excess charges are null. The saturation curves were obtained for IBA CC003 for all studied beam collimation, energies, machines and orientation of the chamber in respect to the beam axis. The dependence of the ion collection efficiency on the field size was investigated and it was evaluated the need of applied ion collection efficiency correction factors for the determination of field output factors.

The differences between ion collection efficiency and polarity effect determined in parallel and perpendicular orientations were investigated, through the evaluation of the ratio of these two factors acquired in both orientations.

The PTW 60019 and IBA Razor were positioned with the stem parallel (parallel orientation) to the beam axis and operated with 0 V.

## 2.3 Equivalent square small field size:

The determination of the square small field size $S_{clin}$ for each beam collimation and energy was performed with Gafchromic EBT3 films, due to its high spatial resolution. EBT3 films from lot 04022001 were used.

For calibration curve purposes, 13 strips of 3.6x20.3 cm$^2$ were irradiated with a dose range between 2 Gy and 5.6 Gy with steps of 0.3 Gy and one strip was left unexposed. A field size of 10x10 cm$^2$ was used in order to expose them with homogeneous doses. For calibration and field size measurements procedures SDD of 90 cm was used and the films were covered with 10 cm thick piece of solid water. For each film measurement, 500 MU were delivered.

An Epson Expression 10000XL (Seiko Epson Corporation, Nagano, Japan) flatbed scanner was used. The scanner was warmed up for at least 30 min before readings. Before acquisition and after long pauses, five empty scans were taken to stabilize the lamp. Scans were made in landscape orientation. Imagens were acquired in transmission mode using a glass compressor plate with 3 to 4 mm thick, over the films. Films were scanned in 48-bit RGB mode, with the colour correction tool not active, with 72 ppp of resolution, and saved as TIFF files.

The calibration curves were created in OmniPro I'mRT software version 1.7 for each energy and applied for the exposed films with different beam collimation.

The field dimensions in each direction (crossline and inline) were obtained from the OmniPro I'mRT software. Using equation (3) and (4), the equivalent square small field size was determined for square and circular small field sizes (IAEA TRS-483 2017), respectively,

$$S_{clin} = \sqrt{AB} \qquad (3)$$
$$S_{clin} = r\sqrt{\pi} = 1.77r \qquad (4)$$

where A and B correspond to the crossline and inline dosimetric widths, defined by FWHM at the measurement depth and $r$ corresponds to the radius of the circular field defined by the points which on average, the dose level amounts to 50% of the maximum dose at the measurement depth.

## 2.4 Monte Carlo

The Monte Carlo simulations were made using the Penelope (NEA 2005) system, to simulate the physical interactions. For the geometry definition, and tracking and scoring of the particles, a homemade system called Ulysses was used (Peralta L and Louro A 2014).

Phase-space files provided by Varian for TruBeam/Edge linear accelerator were used. These files were tailed on a plane located just upstream of the movable jaws and they were used as a radiation source. Downstream, it was created the geometry of the jaws X and Y, base plate, MLC (high definition and Millenium) and water phantom. The IBA CC003 was modeled according to the blueprints provided by the manufacturer, using three spheres, one inside the other (external: graphite; medium: shonka; inner: graphite) (Figure 1).

The simulation was performed for four types of geometry configurations: 1) only jaws; 2) high definition MLC (0.25 cm of thickness for central leafs at isocentre) named MLC1; 3) Millennium MLC (0.5 cm of thickness for central leafs at isocentre) named MLC2; and 4) Cones. The field sizes used were the same as the experimental measurements.

To improve the statistics of the Monte Carlo results 1) the phase-space files were read several times until the defined total number of events was achieved ($2x10^9$) and 2) each program for each geometry configuration was run 20 times with different seeds, to minimize the uncertainty of the simulation result.

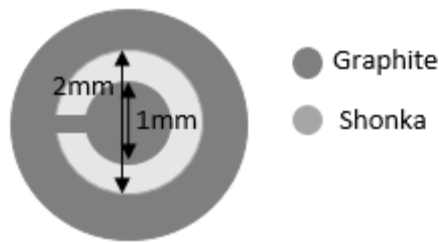

Figure 1 - Dimensions of simulated IBA CC003.

**2.5 Profiles and percentage depth doses (PDDs)**

Beams profiles in crossline and inline directions and PDDs were evaluated for small fields collimated by the MLC (3x3, 2x2 1x1 and 0.5x0.5 cm$^2$) for all energies presented in this study. The evaluations were done using the IBA CC003, PTW60019 and IBA Razor detectors and MC simulation.

The measurements were performed in the water phantom using a scanning resolution of 0.2 mm for beam profiles and 0.5 mm for PDDs. Both detectors were placed in parallel orientation in respect to the beam axis. Profiles were acquired at isocentre with SSD of 90 cm and depth of 10 cm and PDD at SSD of 100 cm.

FWHM and penumbra (distance between 80% and 20% dose) of beam profiles was evaluated and compared between the three detectors and MC simulation. Additionally, for the evaluation of the beam profiles and PDDs, Gamma analysis (Low D et al. 1998) was performed, using the passing criteria of 3%/3 mm and 2%/2 mm.

**2.6 Field Output factors**

For the IBA CC003 the field output factor was determined without correction factors, because this detector is not included in the TRS-483 list of field output correction factors. However, it was applied the polarity correction factor to the raw measurements, thus it was followed the given equation:

$$\Omega^{f_{clin},f_{ref}}_{Q_{clin},Q_{ref}\,IBA\,CC003} = \frac{M^{f_{clin}}_{Q_{clin}}}{M^{f_{ref}}_{Q_{ref}}} \frac{k_{pol_{clin}}}{k_{pol_{ref}}} \quad (5)$$

The differences between field output factors determined in parallel and perpendicular orientations of IBA CC003 were investigated, through the evaluation of the ratio of the field output factors acquired in both orientations.

For PTW60019 detector the field output factors were determined using the following equation:

$$\Omega^{f_{clin},f_{ref}}_{Q_{clin},Q_{ref}\,PTW60019} = \frac{M^{f_{clin}}_{Q_{clin}}}{M^{f_{ref}}_{Q_{ref}}} k^{f_{clin},f_{ref}}_{Q_{clin},Q_{ref}} \quad (6)$$

For the IBA Razor detector the field output factors were determined using the "Daisy-Chainning" method (IAEA TRS-483 2017, Griessbach I et al. 2005). For this method the ratio of the readings is measured between an intermediate and the reference field size using a suitable ionization chamber, and then measured the ratio of the readings between a clinical field size (defined by MLC, jaws or cones) and the intermediate field size using the diode detector. The ionization chamber used was the CC04 from IBA and the intermediate field size ($f_{int}$) was 4x4 cm$^2$. The field output factor used was given by:

$$\Omega^{f_{clin},f_{ref}}_{Q_{clin},Q_{ref}\,IBA\,Razor} = \left[\frac{M^{f_{clin}}_{Q_{clin}} k^{f_{clin},f_{ref}}_{Q_{clin},Q_{ref}}}{M^{f_{4x4}}_{Q_{4x4}} k^{f_{4x4},f_{ref}}_{Q_{4x4},Q_{ref}}}\right]_{Razor} \left[\frac{M^{f_{4x4}}_{Q_{4x4}}}{M^{f_{ref}}_{Q_{ref}}}\right]_{CC04} \quad (7)$$

For each measurement performed using any detector, 200 MU were delivered and at least three measurements of collected charge were taken for each setup.

The field output factors determined using MC simulation were calculated using the following equation:

$$\Omega^{f_{clin},f_{ref}}_{Q_{clin},Q_{ref}\,MC} = \frac{D^{f_{clin}}_{det,Q_{clin}}}{D^{f_{ref}}_{det,Q_{ref}}} \quad (8)$$

here the $D^{f_{clin}}_{det,Q_{clin}}$ corresponds to the deposited energy in the simulated IBA CC003 for a given field size between 4x4 and 0.5x0.5 cm$^2$ for jaws and MLC and between 1.75 and 0.75 cm for cones and $D_{det,Q_{ref}}^{f_{ref}}$ corresponds to the deposited energy in the simulated IBA CC003 for the reference field size 10x10 cm².

The *msr* field was replaced for *ref* field because the 10x10 cm² reference field was used.

## 2.7 Field Output Correction factors

The field output correction factors were derived through three methods: 1) Reference detectors with known output correction factors, 2) Comparison between Monte Carlo simulation and experimental measurements, and 3) Only Monte Carlo Simulation.

For the first method (Experimental method) it was assumed the PTW 60019 and the IBA Razor as the reference detectors and the field output correction factors for IBA CC003 was obtained for parallel and perpendicular orientation using the following equation:

$$k_{Q_{clin},Q_{ref}}^{f_{clin},f_{ref}}[\text{IBA CC003}] = \frac{\Omega_{Q_{clin},Q_{ref}\,ref\,det}^{f_{clin},f_{ref}}}{\Omega_{Q_{clin},Q_{ref}\,IBA\,CC003}^{f_{clin},f_{ref}}} \quad (9)$$

For the second method (Hybrid method) it was assumed the Monte Carlo results as a reference and the field output correction factors for IBA CC003 in parallel orientation was obtained using the following equation:

$$k_{Q_{clin},Q_{ref}}^{f_{clin},f_{ref}}[\text{IBA CC003}] = \frac{\Omega_{Q_{clin},Q_{ref}\,MC}^{f_{clin},f_{ref}}}{\Omega_{Q_{clin},Q_{ref}\,IBA\,CC003_{para}}^{f_{clin},f_{ref}}} \quad (10)$$

For the third method (MC method) the field output correction factors were derived according to the following equation:

$$k_{Q_{clin},Q_{ref}}^{f_{clin},f_{ref}}[\text{IBA CC003}] = \frac{\Omega_{Q_{clin},Q_{ref}\,MC_{water}}^{f_{clin},f_{ref}}}{\Omega_{Q_{clin},Q_{ref}\,MC_{IBA\,CC003}}^{f_{clin},f_{ref}}} \quad (11)$$

where $\Omega_{Q_{clin},Q_{ref}\,MC_{IBA\,CC003}}^{f_{clin},f_{ref}}$ corresponds the field output factor obtained in the simulated IBA CC003 for a given clinical field and $\Omega_{Q_{clin},Q_{ref}\,MC_{water}}^{f_{clin},f_{ref}}$ is given by:

$$\Omega_{Q_{clin},Q_{ref}\,MC_{water}}^{f_{clin},f_{ref}} = \frac{D_{w,Q_{clin}}^{f_{clin}}}{D_{w,Q_{ref}}^{f_{ref}}} \quad (12)$$

where $D_{w,Q_{clin}}^{f_{clin}}$ and $D_{w,Q_{ref}}^{f_{ref}}$ were obtained taking the deposited energy from a sphere of water with a radius of 0.1 cm for the clinical fields and the reference field 10x10 cm² respectively.

## 2.8 Uncertainties

Measurement and Monte Carlo uncertainties were estimated following the recommendations of the Evaluation of measurement data – Guide to expression of uncertainty in measurement (BIPM 2008) and from IAEA publications TECDOC-158529 (IAEA 2008) and TRS-39830 (Andreo P et al. 2004).

The field output uncertainties were calculated for experimental measurements and MC simulations for each machine, energy, beam collimation, detector and orientation of the chamber to respect the beam axis. An uncertainty budget was created and it was evaluated the different uncertainty sources which contribute to the measurement parameters.

For the field output factor acquired by experimental measurements, the sources of uncertainty considered were 1) repeated measurements, 2) calibration certificate of electrometer and 3) resolution of electrometer. For the analysis of repeated measurement a Type A standard uncertainty was determined, in contrast, a Type B standard uncertainty was used to evaluate the influence of the calibration and resolution of the PTW Unidos electrometer. For $k_{Q_{clin},Q_{ref}}^{f_{clin},f_{ref}}$, it was considered a combined uncertainty of 1% for field sizes larger than 1x1cm² and 2% for field sizes equal to or smaller than 1x1 cm².

For the field output factor acquired through MC simulation, the standard uncertainties of $D_{det,Q_{clin}}^{f_{clin}}$ and $D_{det,Q_{ref}}^{f_{ref}}$ were obtained through the standard deviation of the mean of the deposited energy on the simulated IBA CC003 and also 20 runs done for each geometric configuration and energy.

The uncertainties for field output correction factors were also calculated using an uncertainty budget, for each energy, beam collimation (MLC, Jaws and Cones) and field size.

## 3. Results and discussion:

### 3.1 Equivalent square small field sizes

Data for the average of equivalent square field size between the three linear accelerators and corresponding expanded uncertainty are present in Table 1 and Table 2 for field sizes defined by MLC and jaws. In Table 3 it is presented the equivalent square field size and the associated uncertainty, for the cones acquired in the linear accelerator Edge. For the nominal $S_{clin} \geq 1$ cm, the differences found are less than 0.5 mm, except for the energy 10X and the field size defined by the MLC, which presents a difference of 0.8 mm. For $S_{clin} < 1$ cm, the differences found are higher than 1.0 mm and can achieve 1.8 mm, especially for the smallest field size.

Throughout this work, the field sizes will be indicated with nominal values, however, they will represent without exception the corresponding $S_{clin}$ values.

Table 1 - Equivalent square field size ($S_{clin}$) for field sizes defined by the MLC.

| Nominal Field Size (cm) | $S_{clin} \pm U$ (MLC) | | | |
|---|---|---|---|---|
| | 6X | 10X | 6X-FFF | 10X-FFF |
| MLC 10.0x10.0 | 10.01 ± 0.14 | 10.03 ± 0.14 | 9.95 ± 0.14 | 9.89 ± 0.14 |
| MLC 4.0x4.0 | 4.01 ± 0.14 | 4.02 ± 0.14 | 4.02 ± 0.14 | 3.99 ± 0.14 |
| MLC 3.0x3.0 | 3.03 ± 0.14 | 3.04 ± 0.14 | 3.03 ± 0.14 | 3.02 ± 0.14 |
| MLC 2.0x2.0 | 2.02 ± 0.14 | 2.03 ± 0.14 | 2.02 ± 0.14 | 2.02 ± 0.14 |
| MLC 1.0x1.0 | 1.05 ± 0.14 | 1.08 ± 0.14 | 1.05 ± 0.14 | 1.05 ± 0.14 |
| MLC 0.5x0.5 | 0.60 ± 0.14 | 0.68 ± 0.14 | 0.58 ± 0.14 | 0.62 ± 0.14 |

Table 2 - Equivalent square field size ($S_{clin}$) for field sizes defined by the Jaws.

| Nominal Field Size (cm) | $S_{clin} \pm U$ (Jaws) | | | |
|---|---|---|---|---|
| | 6X | 10X | 6X-FFF | 10X-FFF |
| 10.0x10.0 | 9.99 ± 0.15 | 9.96 ± 0.17 | 9.91 ± 0.15 | 9.84 ± 0.15 |
| 4.0x4.0 | 3.96 ± 0.15 | 3.94 ± 0.16 | 3.95 ± 0.15 | 3.93 ± 0.14 |
| 3.0x3.0 | 2.99 ± 0.15 | 2.95 ± 0.16 | 2.97 ± 0.15 | 2.96 ± 0.15 |
| 2.0x2.0 | 1.96 ± 0.15 | 1.94 ± 0.17 | 1.94 ± 0.16 | 1.95 ± 0.15 |
| 1.5x1.5 | 1.45 ± 0.15 | 1.46 ± 0.15 | 1.45 ± 0.15 | 1.46 ± 0.15 |
| 1.0x1.0 | 0.99 ± 0.14 | 1.02 ± 0.16 | 0.97 ± 0.15 | 0.99 ± 0.14 |
| 0.8x0.8 | 0.82 ± 0.14 | 0.86 ± 0.16 | 0.79 ± 0.15 | 0.83 ± 0.14 |
| 0.5x0.5 | 0.60 ± 0.14 | 0.65 ± 0.15 | 0.57 ± 0.15 | 0.62 ± 0.14 |

Table 3 - Equivalent square field size ($S_{clin}$) for field sizes defined by the Cones.

| Nominal Field Size (cm) | Sclin ± U (Cones) | | |
|---|---|---|---|
| | Energy | | |
| | 6X | 6X-FFF | 10X-FFF |
| Cone 1.75 | 1.54 ± 0.14 | 1.54 ± 0.14 | 1.53 ± 0.14 |
| Cone 1.5 | 1.33 ± 0.14 | 1.33 ± 0.14 | 1.32 ± 0.14 |
| Cone 1.25 | 1.10 ± 0.14 | 1.12 ± 0.14 | 1.10 ± 0.14 |
| Cone 1.0 | 0.90 ± 0.14 | 0.89 ± 0.14 | 0.88 ± 0.14 |
| Cone 7.5 | 0.69 ± 0.14 | 0.67 ± 0.14 | 0.68 ± 0.14 |

### 3.2 Profiles and PDD

It is observed the differences of PDDs (between detectors and detectors and simulation) increase with the depth and the decrease of the field size, however the majority of the differences are in the buildup region. The PDDs for IBA CC003 present a slightly overestimated dose with increasing depths when compared with IBA Razor and PTW 60019, with the decreasing of the field size, as reported by Reggiori (Reggiori G et al. 2017). The Gamma Analysis show good results (> 90%) for field sizes above 1x1 cm$^2$, for the two passing criteria and for the comparison between detectors, however between detectors and simulation only for the criteria 3%/3 mm, the Gamma results are above 90% (Table 4, 5, 6 and 7). For field size 0.5x0.5 cm$^2$ and evaluation between detectors, the worst Gamma results were between the detectors IBA CC003 and IBA Razor, maybe because Razor diode lead to an overestimation of the low energies (kV energy range) in the build-up region, and since PDD are normalized to $d_{max}$, this effect induces the under response of the detector in depth. Gamma results between measurements and simulation for the field size 0.5x0.5 cm$^2$ were not satisfactory, because the number of particles, resulting from the phase-space files, is too few for this field size, which is something that is difficult to resolve because we are dependent on the limited number of particle existing in the phase-space file provided by Varian.

Table 4 - Gamma analysis results for passing criteria of 3%/3 mm and 2%/2 mm for the PDD acquired with the three detectors and MC for 6X.

| Nominal Field Size (cm) | Gamma Analysis | IBA CC003 vs. PTW60019 | IBA CC003 vs. IBA Razor | PTW60019 vs. IBA Razor | IBA CC003 vs. MC | PTW60019 vs. MC | IBA Razor vs. MC |
|---|---|---|---|---|---|---|---|
| MLC 3.0x3.0 | 2%/2mm | 99.42% | 94.03% | 91.79% | 96.25% | 88.20% | 99.21% |
| | 3%/3mm | 100.00% | 95.52% | 95.52% | 98.75% | 99.38% | 99.21% |
| MLC 2.0x2.0 | 2%/2mm | 98.22% | 94.03% | 96.26% | 91.88% | 96.25% | 96.00% |
| | 3%/3mm | 99.41% | 97.01% | 98.51% | 98.12% | 98.12% | 98.40% |
| MLC 1.0x1.0 | 2%/2mm | 99.41% | 97.01% | 97.01% | 98.14% | 96.89% | 94.44% |
| | 3%/3mm | 99.41% | 99.25% | 97.76% | 98.76% | 98.76% | 97.62% |
| MLC 0.5x0.5 | 2%/2mm | 95.21% | 22.23% | 97.73% | 8.12% | 6.25% | 6.40% |
| | 3%/3mm | 99.40% | 96.97% | 99.24% | 12.50% | 8.12% | 9.60% |

Table 5 - Gamma analysis results for passing criteria of 3%/3 mm and 2%/2 mm for the PDD acquired with the three detectors and MC for 10X.

| Nominal Field Size (cm) | Gamma Analysis | IBA CC003 vs. PTW60019 | IBA CC003 vs. IBA Razor | PTW60019 vs. IBA Razor | IBA CC003 vs. MC | PTW60019 vs. MC | IBA Razor vs. MC |
|---|---|---|---|---|---|---|---|
| MLC 3.0x3.0 | 2%/2mm | 94.29% | 93.53% | 100.00% | 94.41% | 96.89% | 96.00% |
|  | 3%/3mm | 94.86% | 97.12% | 100.00% | 98.14% | 99.38% | 99.20% |
| MLC 2.0x2.0 | 2%/2mm | 93.14% | 90.71% | 97.86% | 90.62% | 97.50% | 97.60% |
|  | 3%/3mm | 97.71% | 95.00% | 100.00% | 97.50% | 98.75% | 99.20% |
| MLC 1.0x1.0 | 2%/2mm | 95.38% | 91.18% | 100.00% | 82.50% | 86.25% | 88.80% |
|  | 3%/3mm | 98.84% | 96.32% | 100.00% | 93.75% | 97.50% | 96.00% |
| MLC 0.5x0.5 | 2%/2mm | 20.71% | 18.66% | 99.25% | 16.88% | 13.12% | 18.40% |
|  | 3%/3mm | 80.47% | 41.04% | 99.25% | 21.88% | 17.50% | 20.00% |

Table 6 - Gamma analysis results for passing criteria of 3%/3 mm and 2%/2 mm for the PDD acquired with the three detectors and MC for 6X-FFF.

| Nominal Field Size (cm) | Gamma Analysis | IBA CC003 vs. PTW60019 | IBA CC003 vs. IBA Razor | PTW60019 vs. IBA Razor | IBA CC003 vs. MC | PTW60019 vs. MC | IBA Razor vs. MC |
|---|---|---|---|---|---|---|---|
| MLC 3.0x3.0 | 2%/2mm | 98.22% | 94.78% | 96.27% | 98.14% | 98.14% | 98.41% |
|  | 3%/3mm | 99.41% | 97.01% | 97.76% | 98.76% | 98.76% | 99.21% |
| MLC 2.0x2.0 | 2%/2mm | 98.22% | 93.28% | 97.76% | 58.39% | 75.78% | 91.27% |
|  | 3%/3mm | 99.41% | 97.01% | 97.76% | 95.03% | 95.03% | 99.21% |
| MLC 1.0x1.0 | 2%/2mm | 99.40% | 95.52% | 96.21% | 93.12% | 96.25% | 96.80% |
|  | 3%/3mm | 99.40% | 97.76% | 97.73% | 98.75% | 98.12% | 98.40% |
| MLC 0.5x0.5 | 2%/2mm | 98.20% | 21.97% | 96.21% | 5.62% | 5.62% | 5.60% |
|  | 3%/3mm | 99.40% | 97.73% | 97.73% | 9.38% | 7.50% | 8.00% |

Table 7 - Gamma analysis results for passing criteria of 3%/3 mm and 2%/2 mm for the PDD acquired with the three detectors and MC for 10X-FFF.

| Nominal Field Size (cm) | Gamma Analysis | IBA CC003 vs. PTW60019 | IBA CC003 vs. IBA Razor | PTW60019 vs. IBA Razor | IBA CC003 vs. MC | PTW60019 vs. MC | IBA Razor vs. MC |
|---|---|---|---|---|---|---|---|
| MLC 3.0x3.0 | 2%/2mm | 96.00% | 93.57% | 96.43% | 96.89% | 98.14% | 99.21% |
|  | 3%/3mm | 99.43% | 97.14% | 97.86% | 98.76% | 98.76% | 99.21% |
| MLC 2.0x2.0 | 2%/2mm | 94.29% | 93.48% | 96.38% | 96.27% | 96.89% | 99.20% |
|  | 3%/3mm | 97.71% | 97.10% | 98.55% | 97.52% | 98.76% | 99.20% |
| MLC 1.0x1.0 | 2%/2mm | 97.66% | 93.38% | 97.79% | 96.88% | 93.75% | 67.20% |
|  | 3%/3mm | 99.42% | 97.06% | 98.53% | 97.50% | 97.50% | 96.80% |
| MLC 0.5x0.5 | 2%/2mm | 87.57% | 24.63% | 97.76% | 8.12% | 5.62% | 7.20% |
|  | 3%/3mm | 95.86% | 95.52% | 98.51% | 15.62% | 9.38% | 12.00% |

No significant differences were found for FWHM and $S_{clin}$ between the three detectors used, except for the smallest field size 0.5x0.5 cm$^2$, which presents higher $S_{clin}$ values for IBA CC003 compared to PTW60019 and IBA Razor (maximum difference of 0.8 mm). For the field size 1x1 cm$^2$, TrueBeam2 presents higher $S_{clin}$ values compared to Edge and TrueBeam1, which can be related to the thickness of the MLC leafs of the TrueBeam2, which are thicker than the ones of Edge and TrueBeam1. Comparing $S_{clin}$ values between the three detectors and EBT3 film, it was found higher differences for the field size 0.5x0.5 cm$^2$, but on average less than 0.5 mm.

For the profiles, Gamma analysis show that differences between the detectors and detectors and simulation increase with the decreasing of the field size (Table 8, 9, 10 and 11).  Only for the passing criteria of 3%/3 mm were the results satisfactory (> 90%) above 2x2 cm$^2$ field size and in some cases for the field size 1x1 cm$^2$. For the evaluation between detectors, higher differences are between the detectors IBA CC003 and IBA Razor, which is explained by the difference of the active volumes of these two detectors. For the evaluation between detectors and simulation, the Gamma analysis results for profiles are not as satisfactory as for PDD, possibly because the profiles present much less points to evaluate. This is a big disadvantage for the small field especially on the penumbra, because even two points considered very close can be far enough to have a large difference in % dose, and this influences the Gamma analysis result.

Table 8 - Gamma analysis results for passing criteria of 3%/3 mm and 2%/2 mm for the Profiles acquired with the three detectors and MC for 6X.

| Nominal Field Size (cm) | Gamma Analysis | IBA CC003 vs. PTW60019 | IBA CC003 vs. IBA Razor | PTW60019 vs. IBA Razor | IBA CC003 vs. MC | PTW60019 vs. MC | IBA Razor vs. MC |
|---|---|---|---|---|---|---|---|
| MLC 3.0x3.0 | 2%/2mm | 84.95% | 85.56% | 92.22% | 81.82% | 79.69% | 76.92% |
|  | 3%/3mm | 90.32% | 93.33% | 96.67% | 87.88% | 90.62% | 87.69% |
| MLC 2.0x2.0 | 2%/2mm | 85.88% | 62.82% | 84.62% | 67.86% | 70.37% | 67.92% |
|  | 3%/3mm | 92.94% | 82.05% | 96.15% | 78.56% | 81.48% | 83.02% |
| MLC 1.0x1.0 | 2%/2mm | 81.16% | 56.45% | 87.10% | 75.56% | 75.00% | 63.64% |
|  | 3%/3mm | 86.96% | 80.65% | 95.16% | 86.67% | 84.09% | 77.27% |
| MLC 0.5x0.5 | 2%/2mm | 45.61% | 51.79% | 66.07% | 65.00% | 50.00% | 40.00% |
|  | 3%/3mm | 75.44% | 66.07% | 78.57% | 75.00% | 70.00% | 57.50% |

Table 9 - Gamma analysis results for passing criteria of 3%/3 mm and 2%/2 mm for the Profiles acquired with the three detectors and MC for 10X.

| Nominal Field Size (cm) | Gamma Analysis | IBA CC003 vs. PTW60019 | IBA CC003 vs. IBA Razor | PTW60019 vs. IBA Razor | IBA CC003 vs. MC | PTW60019 vs. MC | IBA Razor vs. MC |
|---|---|---|---|---|---|---|---|
| MLC 3.0x3.0 | 2%/2mm | 86.73% | 89.13% | 92.39% | 83.33% | 78.79% | 77.27% |
|  | 3%/3mm | 92.86% | 92.39% | 97.83% | 89.39% | 92.42% | 87.88% |
| MLC 2.0x2.0 | 2%/2mm | 84.27% | 81.93% | 92.50% | 67.24% | 70.18% | 71.93% |
|  | 3%/3mm | 91.01% | 93.98% | 95.00% | 87.93% | 87.72% | 87.72% |
| MLC 1.0x1.0 | 2%/2mm | 78.08% | 61.43% | 83.58% | 82.22% | 82.22% | 68.89% |
|  | 3%/3mm | 90.41% | 81.43% | 91.04% | 88.89% | 84.44% | 88.89% |
| MLC 0.5x0.5 | 2%/2mm | 71.43% | 46.43% | 58.93% | 72.50% | 67.50% | 42.50% |
|  | 3%/3mm | 80.36% | 64.29% | 80.36% | 77.50% | 87.50% | 62.50% |

Table 10 - Gamma analysis results for passing criteria of 3%/3 mm and 2%/2 mm for the Profiles acquired with the three detectors and MC for 6X-FFF.

| Profiles 6X-FFF | | | | | | | |
|---|---|---|---|---|---|---|---|
| Nominal Field Size (cm) | Gamma Analysis | IBA CC003 vs. PTW60019 | IBA CC003 vs. IBA Razor | PTW60019 vs. IBA Razor | IBA CC003 vs. MC | PTW60019 vs. MC | IBA Razor vs. MC |
| MLC 3.0x3.0 | 2%/2mm | 85.71% | 84.71% | 80.95% | 69.23% | 72.31% | 72.31% |
|  | 3%/3mm | 91.21% | 91.76% | 86.90% | 84.62% | 84.62% | 83.08% |
| MLC 2.0x2.0 | 2%/2mm | 80.52% | 83.56% | 83.56% | 83.64% | 85.19% | 87.04% |
|  | 3%/3mm | 90.91% | 89.04% | 91.78% | 89.09% | 92.59% | 94.44% |
| MLC 1.0x1.0 | 2%/2mm | 75.76% | 62.30% | 71.43% | 71.74% | 80.43% | 54.55% |
|  | 3%/3mm | 83.33% | 81.97% | 85.71% | 82.61% | 86.96% | 77.27% |
| MLC 0.5x0.5 | 2%/2mm | 79.63% | 62.50% | 71.15% | 71.79% | 58.97% | 44.74% |
|  | 3%/3mm | 83.33% | 82.14% | 82.69% | 79.49% | 84.62% | 57.89% |

Table 11 - Gamma analysis results for passing criteria of 3%/3 mm and 2%/2 mm for the Profiles acquired with the three detectors and MC for 10X-FFF.

| Profiles 10X-FFF | | | | | | | |
|---|---|---|---|---|---|---|---|
| Nominal Field Size (cm) | Gamma Analysis | IBA CC003 vs. PTW60019 | IBA CC003 vs. IBA Razor | PTW60019 vs. IBA Razor | IBA CC003 vs. MC | PTW60019 vs. MC | IBA Razor vs. MC |
| MLC 3.0x3.0 | 2%/2mm | 90.29% | 88.78% | 90.43% | 83.82% | 85.07% | 86.57% |
|  | 3%/3mm | 95.15% | 88.78% | 94.68% | 91.18% | 92.54% | 91.04% |
| MLC 2.0x2.0 | 2%/2mm | 83.53% | 83.33% | 86.90% | 84.48% | 79.31% | 80.70% |
|  | 3%/3mm | 90.59% | 86.90% | 92.86% | 93.10% | 86.21% | 91.23% |
| MLC 1.0x1.0 | 2%/2mm | 85.92% | 73.13% | 80.60% | 68.89% | 75.56% | 51.11% |
|  | 3%/3mm | 90.14% | 85.07% | 94.03% | 82.22% | 86.67% | 77.78% |
| MLC 0.5x0.5 | 2%/2mm | 70.18% | 56.36% | 71.10% | 77.50% | 72.50% | 50.00% |
|  | 3%/3mm | 84.21% | 78.18% | 86.79% | 85.00% | 85.00% | 60.00% |

**3.3 Polarity correction factor and Ion collection efficiency of IBA CC003**

A strong variation of polarity effect with the field size is observed for all the energies presented in this study in the parallel orientation of the chamber (Figure 2). The polarity effect increases continuously with the decrease of the field size, which is also observed by Looe (Looe et al. 2018) and Reggiori (Reggiori G et al. 2017). Although, for perpendicular orientation of the chamber, the field size dependence is not as relevant, as reported by Looe (Looe et al. 2018) and Gul (Gul et al. 2020). A difference of around 10% and 1.5% was found between the reference field size and the smallest field size for all energies, for parallel and perpendicular orientation of the chamber, respectively.

Assuming the condition $\frac{[k_{pol}]_{para}}{[k_{pol}]_{perp}} - 1 > U(k_{pol})$, we can verify that the polarity effect is significantly different between the two orientations of the chamber within 95% confidence limits, as the field size decreases. The values found for the condition $\frac{[k_{pol}]_{para}}{[k_{pol}]_{perp}} - 1$ were, in most of the cases, above $U(k_{pol})$ and can achieve 0.01 for the smallest field size of 0.5x0.5 cm$^2$.

Small differences in the polarity correction factor were found between the different linear accelerators (<1%). At reference field size, the polarity correction factor for parallel orientation of the chamber was 1.004±0.002, 1.013±0.002, 0.999±0.002 and 1.009±0.002 for 6X, 10X, 6X-FFF and 10X-FFF, respectively. And for perpendicular orientation were 1.012±0.002, 1.019±0.002, 1.008±0.002 and 1.017±0.002 for 6X, 10X, 6X-FFF and 10X-FFF, respectively.

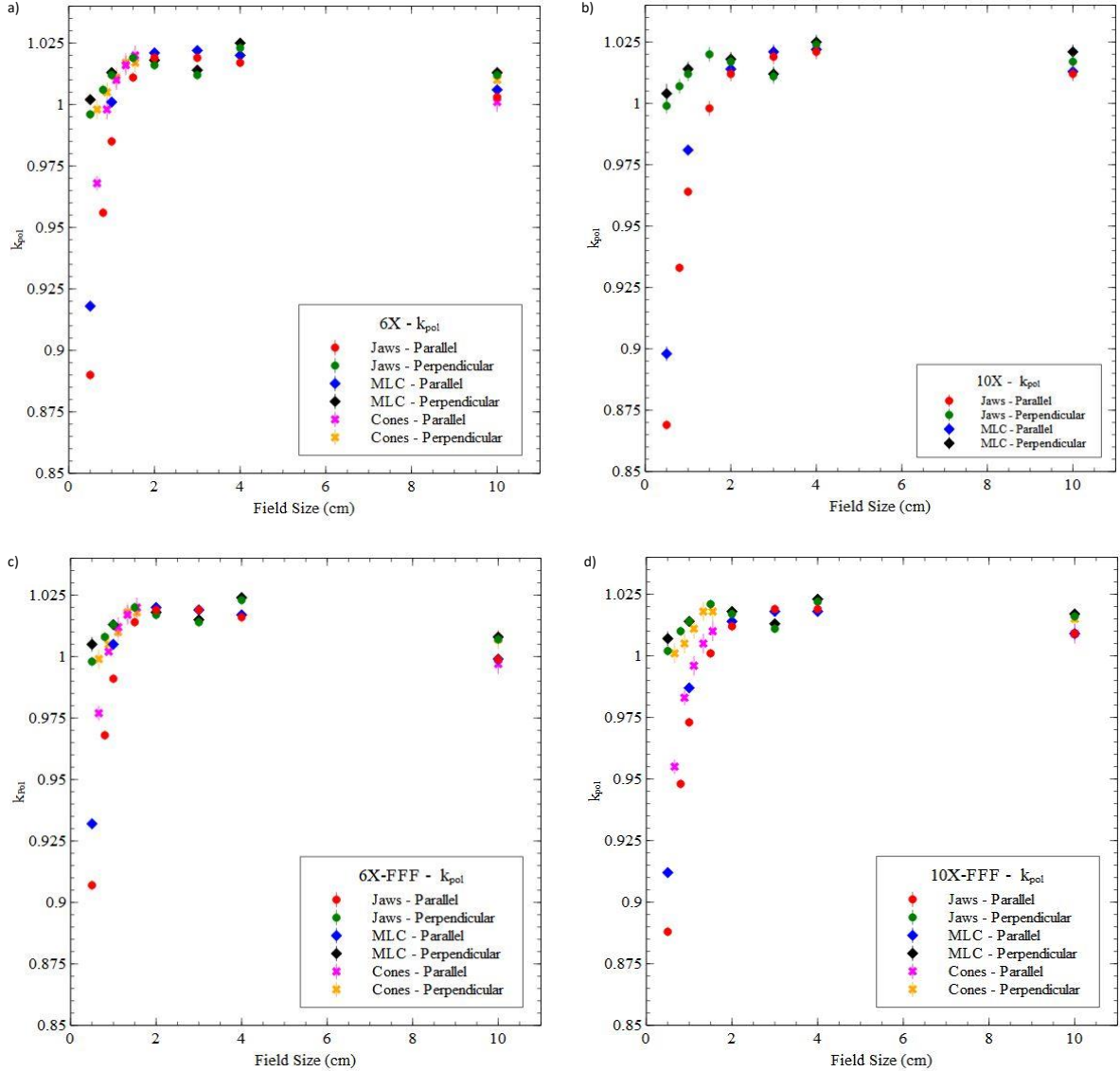

Figure 2 - Average polarity correction factor between the three linear accelerator for field sizes defined by the Jaws, MLC and Cones in parallel and perpendicular orientation of the IBA CC003: a) 6X, b) 10X, c) 6X-FFF and d) 10X-FFF.

Analysis of the ion collection efficiency correction factor with field size (Figure 3) showed no significant variation for the parallel orientation of the chamber, the differences between the reference field size and the small field sizes are less than 0.5%. However for the perpendicular orientation of the chamber it is possible to observe a significant difference between the reference field size and the smallest field size (0.5x0.5 cm$^2$) defined by the jaws for all energies presented in this study, which is around 3%, but this behavior is not observed for the smallest field size defined by the MLC (difference less than 0.5%). Analyzing the ratio between the ion collection efficiency correction factor acquired in both directions, there were differences less than 1%, except for the smallest field size defined by the jaws, where $\frac{[J_s]_{para}}{[J_s]_{perp}} - 1$ is higher than an expanded uncertainty of $[J_s]_{para}$ and $[J_s]_{perp}$, considering $k = 2$. Small differences in the ion collection efficiency correction factor were found between the different linear accelerators (<0.5%). At reference field size, the ion collection efficiency correction factor for parallel

orientation of the chamber was 1.001±0.006, 1.003±0.008, 1.003±0.006 and 1.004±0.006 for 6X, 10X, 6X-FFF and 10X-FFF, respectively. For perpendicular orientation were 1.003±0.006, 1.004±0.008, 1.005±0.006 and 1.005±0.006 for 6X, 10X, 6X-FFF and 10X-FFF, respectively. It was decided not to apply the ion collection efficiency correction factor to determine the field output factor for both orientations of the IBA CC003, because no significant dependence was observed as function of the field size.

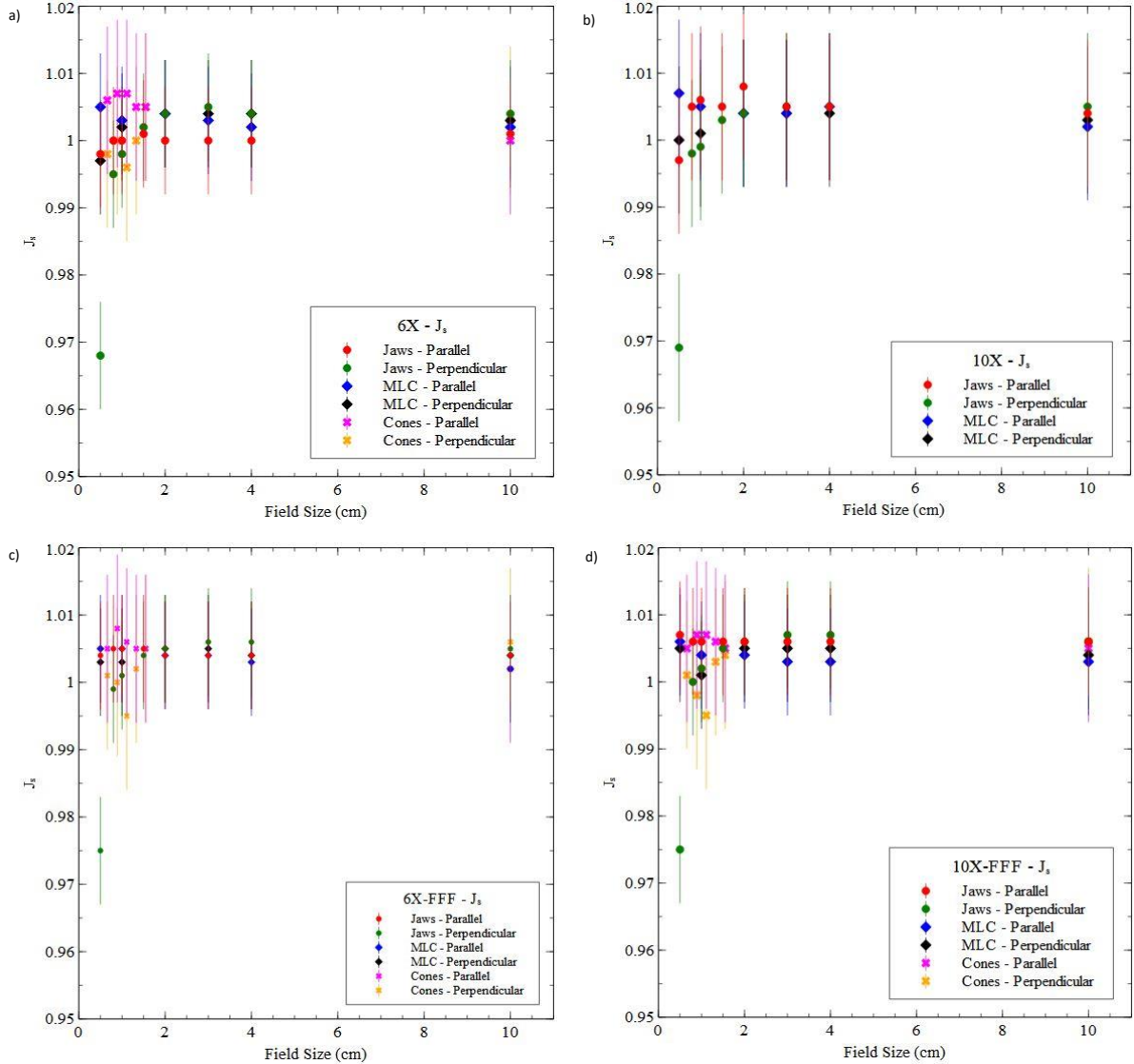

Figure 3- Average ion collection efficiency correction factor between the three linear accelerator for field sizes defined by the Jaws, MLC and Cones in parallel and perpendicular orientation of the IBA CC003: a) 6X, b) 10X, c) 6X-FFF and d) 10X-FFF.

### 3.4 Field Output factors

No significant differences were observed for the OF acquired with IBA CC003 between measurements and Monte Carlo simulation down to the field size 2x2 cm$^2$ defined by the Jaws and MLC, however from field size 1x1 cm$^2$ the field output factor differ significantly between measurements and simulation within 95% confidence limits, because $\left[\Omega_{Q_{clin},Q_{ref}\ IBA\ C003}^{f_{clin},f_{ref}}\right]_{meas} \Big/ \left[\Omega_{Q_{clin},Q_{ref}\ IBA\ C003}^{f_{clin},f_{ref}}\right]_{MC} - 1$ is higher than the expanded uncertainty for the IBA CC003. The same happens for the measurements of OF between parallel and perpendicular orientation of the IBA CC003. For the field sizes defined by Cones, the differences appear for the Cone 7.5 ($S_{clin} = 0.66$ cm) when it is compared measurements versus simulation, and from the Cone 12.5 ($S_{clin} = 1.11$ cm), when it is compared the measurements between parallel and perpendicular orientation of IBA CC003 (Figure 4, 5 and 6).

The measurements OF presented lower values compared to the Monte Carlo OF, because of the volume effect which appears for $S_{clin} < 1.0$ cm, since the dose measured is underestimated for these field sizes.

Small differences in the field output factors were found between the different linear accelerators for all the field sizes (< 1%), except for the field sizes defined by jaws below the field size 0.8x0.8 cm$^2$. It is also observed that for the field sizes defined by the Jaws, the TrueBeam2 presents lower field output values than the others machines for field sizes below 1x1 cm$^2$. This effect is understood due to the field size dimension defined by the Jaws of TrueBeam2 being smaller compared to TrueBeam1 and Edge.

It was found expanded uncertainties less than 1% for measured OF and between 2-3% for simulated OF.

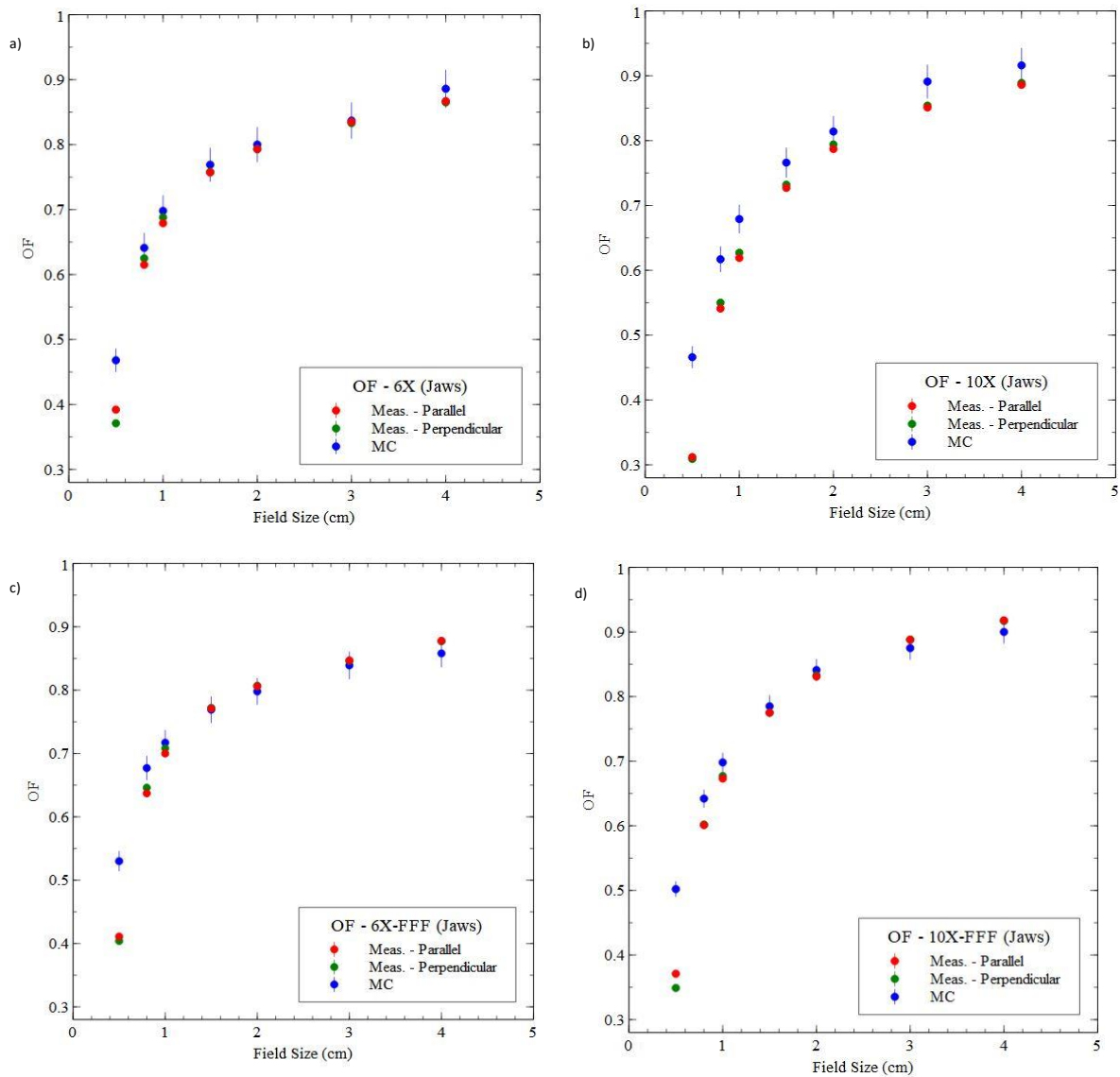

Figure 4 - OF for the beam collimation defined by the Jaws acquired through measurements (in parallel and perpendicular orientation of IBA CC003) and Monte Carlo simulation: a) 6X, b) 10X, c) 6X-FFF and d) 10X-FFF.

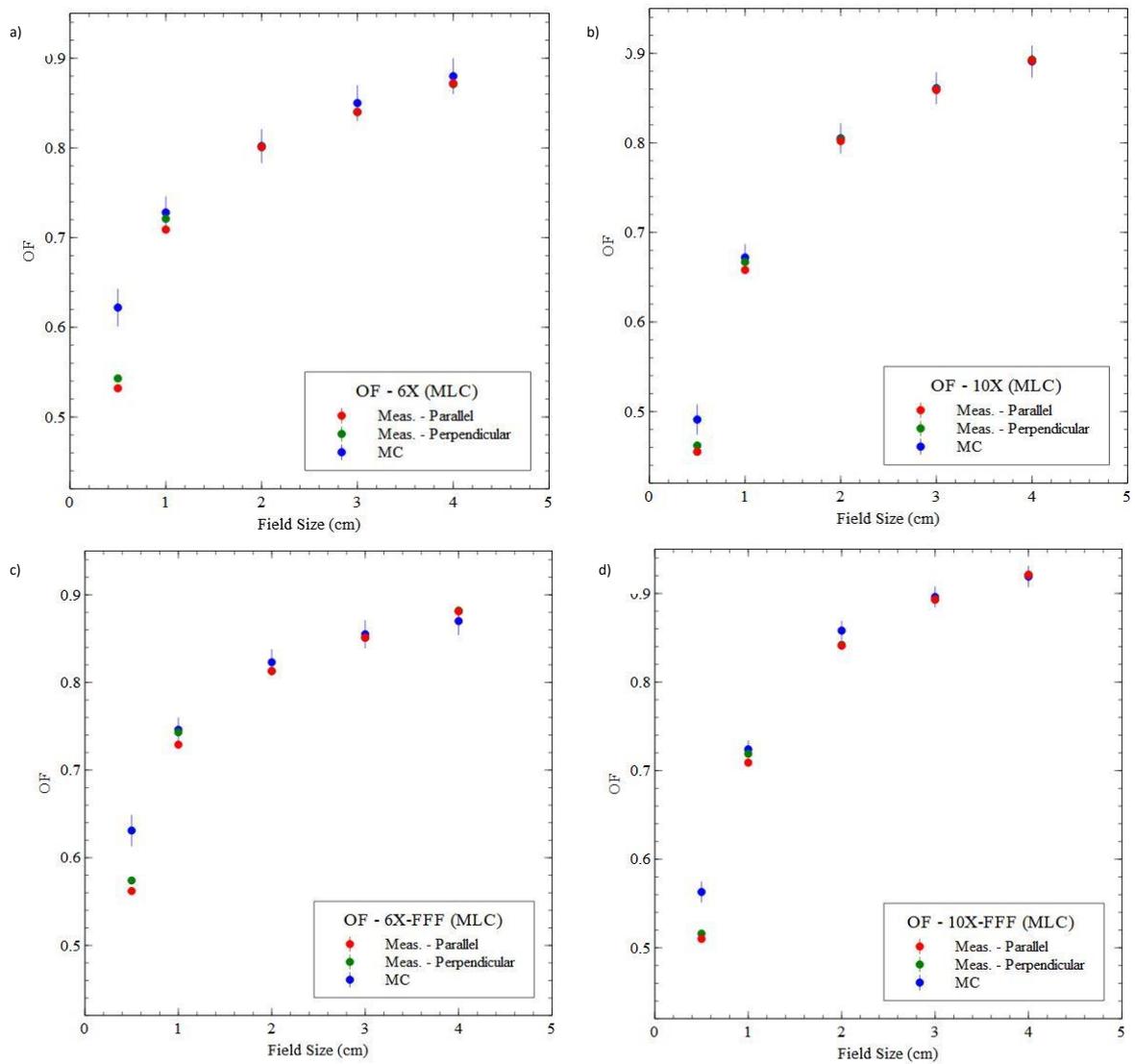

Figure 5 - OF for the beam collimation defined by the MLC acquired through measurements (in parallel and perpendicular orientation of IBA CC003) and Monte Carlo simulation: a) 6X, b) 10X, c) 6X-FFF and d) 10X-FFF.

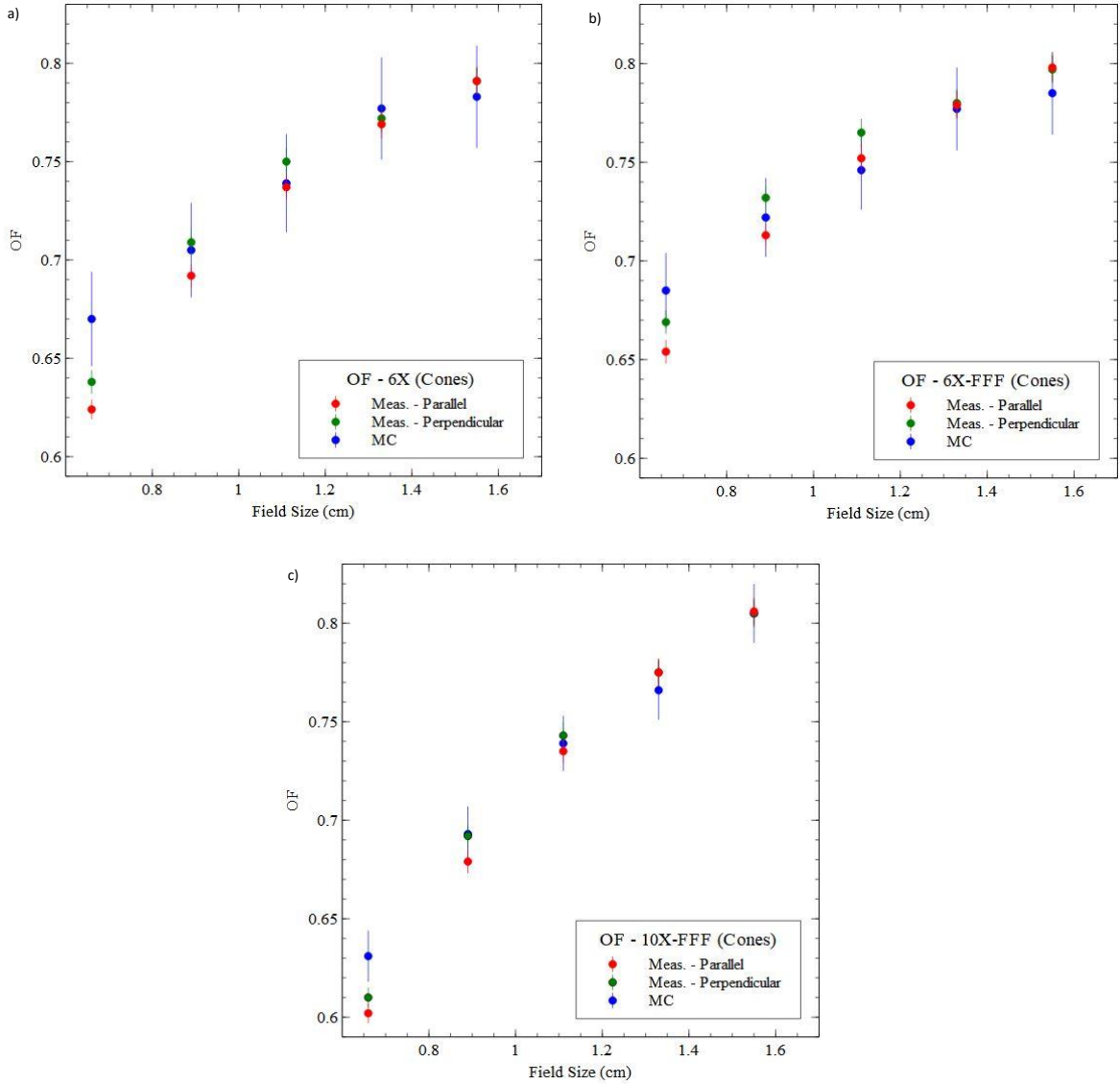

Figure 6 - OF for the beam collimation defined by the Cones acquired through measurements (in parallel and perpendicular orientation of IBA CC003) and Monte Carlo simulation: a) 6X, b) 6X-FFF and c) 10X-FFF.

### 3.5 Field output correction Factors

Analyzing the IBA CC003 field output correction factors obtained through the first (Figure 7 and 8) and second (Figure 9) method, it seems that $k_{Q_{clin},Q_{ref}}^{f_{clin},f_{ref}}$ are needed for $S_{clin} \leq 1$ cm for the three types of beam collimation and beam energy. The value of $k_{Q_{clin},Q_{ref}}^{f_{clin},f_{ref}}$ increases with the decreasing of the field size, which is related to the influence of the volume effect of the IBA CC003 for $S_{clin} < 1$ cm.

$k_{Q_{clin},Q_{ref}}^{f_{clin},f_{ref}}$ obtained with IBA CC003 in parallel orientation (Figure 7) presents a more linear behavior compared with the IBA CC003 in perpendicular orientation (Figure 8). Comparing the results of the IBA CC003 field output correction factors using PTW 60019 and Razor as reference detectors, it is observed that for all field sizes defined by the MLC and for field sizes defined by the Jaws and Cones, for $S_{clin} \geq 0.8$ cm, there is no significant differences in the $k_{Q_{clin},Q_{ref}}^{f_{clin},f_{ref}}$ obtained with the two detectors used as reference (less than 1.0%). However, for the smallest field size defined by the Cones (Cone 0.75) and by the Jaws (0.5x0.5 cm$^2$), higher differences are observed: maximum differences of around 1.5 % for field size defined by the Cones and 5.5 % for the field size defined by the jaws.

The $k_{Q_{clin},Q_{ref}}^{f_{clin},f_{ref}}$ results obtained using the second method (Figure 9), present higher values compared to the first method, which could be related with the limited number of particles existing in the file provided by Varian that limits the number of particles that reach the detector. We have tried to solve this problem, reading the file several times and running the program for each beam configuration 20 times with different seeds, to minimize the uncertainty of the simulation result. Analyzing the IBA CC003 $k_{Q_{clin},Q_{ref}}^{f_{clin},f_{ref}}$ obtained through the third method (Figure 10), it is observed that the field output correction factors decrease with the decreasing of the field size. In this case there is no influence of the volume effect because the volume of the simulated IBA CC003 is the same as the water sphere. Therefore, the effect of perturbation of the particles is more predominant, which can explain the decrease of field output correction factors with the field size.

The expanded uncertainties of $k_{Q_{clin},Q_{ref}}^{f_{clin},f_{ref}}$ for each method were: 1) around 1% - 3% for MLC and Jaws and 2% - 4% for Cones, for the experimental method. For the field sizes down $S_{clin}$ of 1 cm the field output correction factors present an expanded uncertainty higher than the others field sizes, because it was assumed a combined uncertainty of 2% for field sizes less than 1 cm and 1% for field sizes higher than 1 cm for the PTW 60019 and Razor field output correction factors; 2) around 2% - 4% for all the beam collimation for hybrid method; and 3) around 2% - 6% for all the beam collimation for MC method.

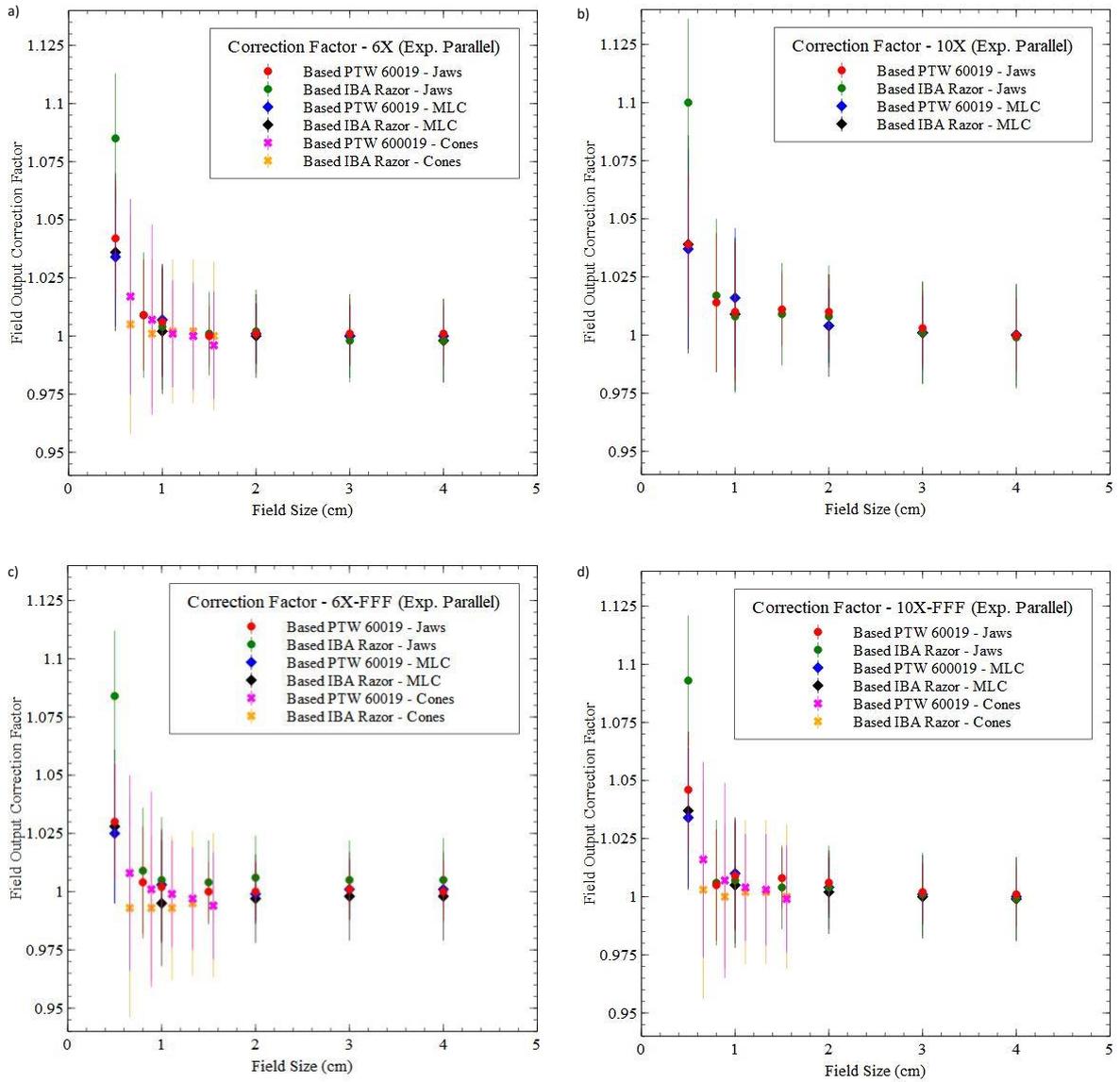

Figure 7 - $k_{Q_{clin},Q_{ref}}^{f_{clin},f_{ref}}$ for all the beam collimation acquired through the Experimental Method using IBA CC003 in parallel orientation: a) 6X, b) 10X, c) 6X-FFF and d) 10X-FFF.

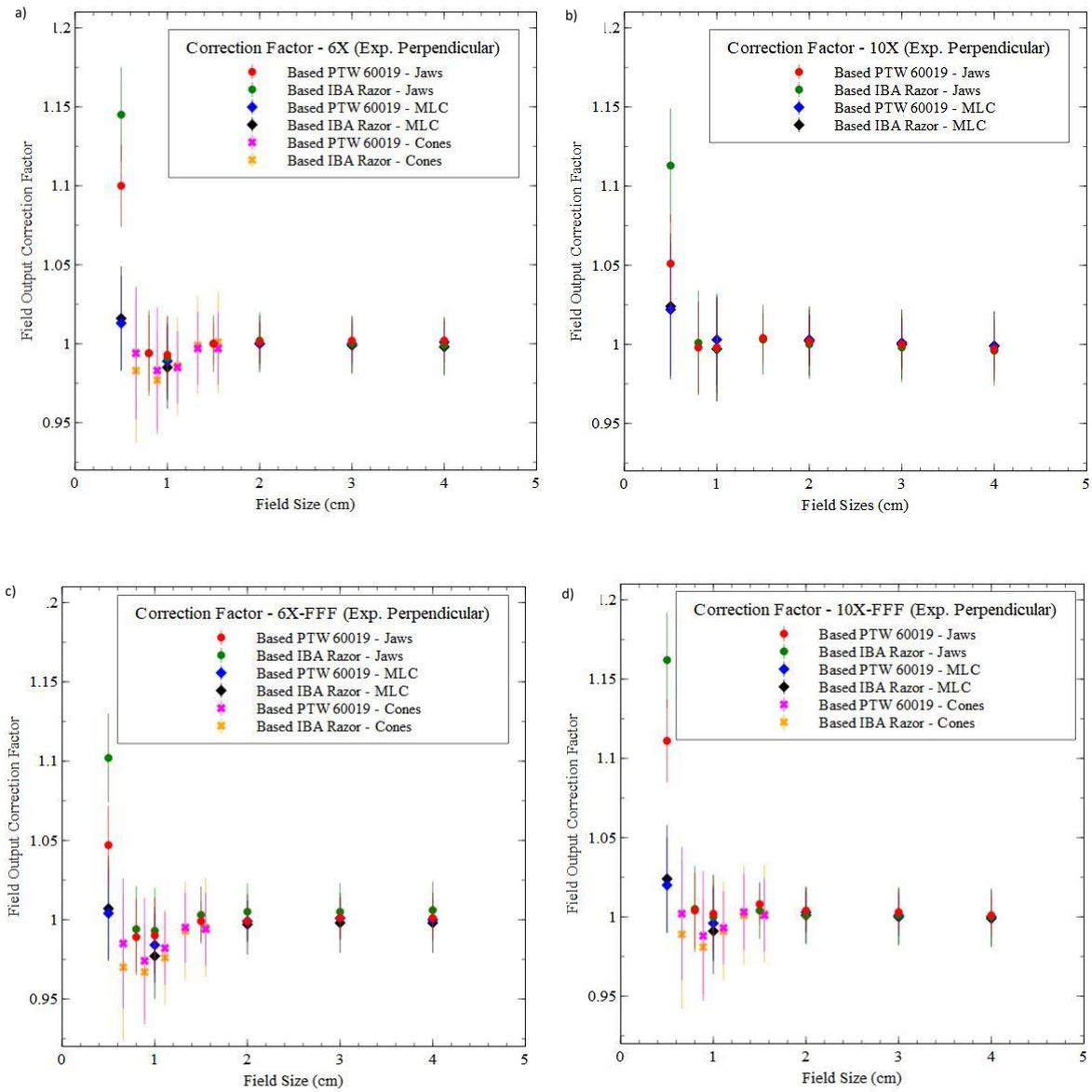

Figure 8 - $k_{Q_{clin},Q_{ref}}^{f_{clin},f_{ref}}$ for all the beam collimation acquired through the Experimental Method using IBA CC003 in perpendicular orientation: a) 6X, b) 10X, c) 6X-FFF and d) 10X-FFF.

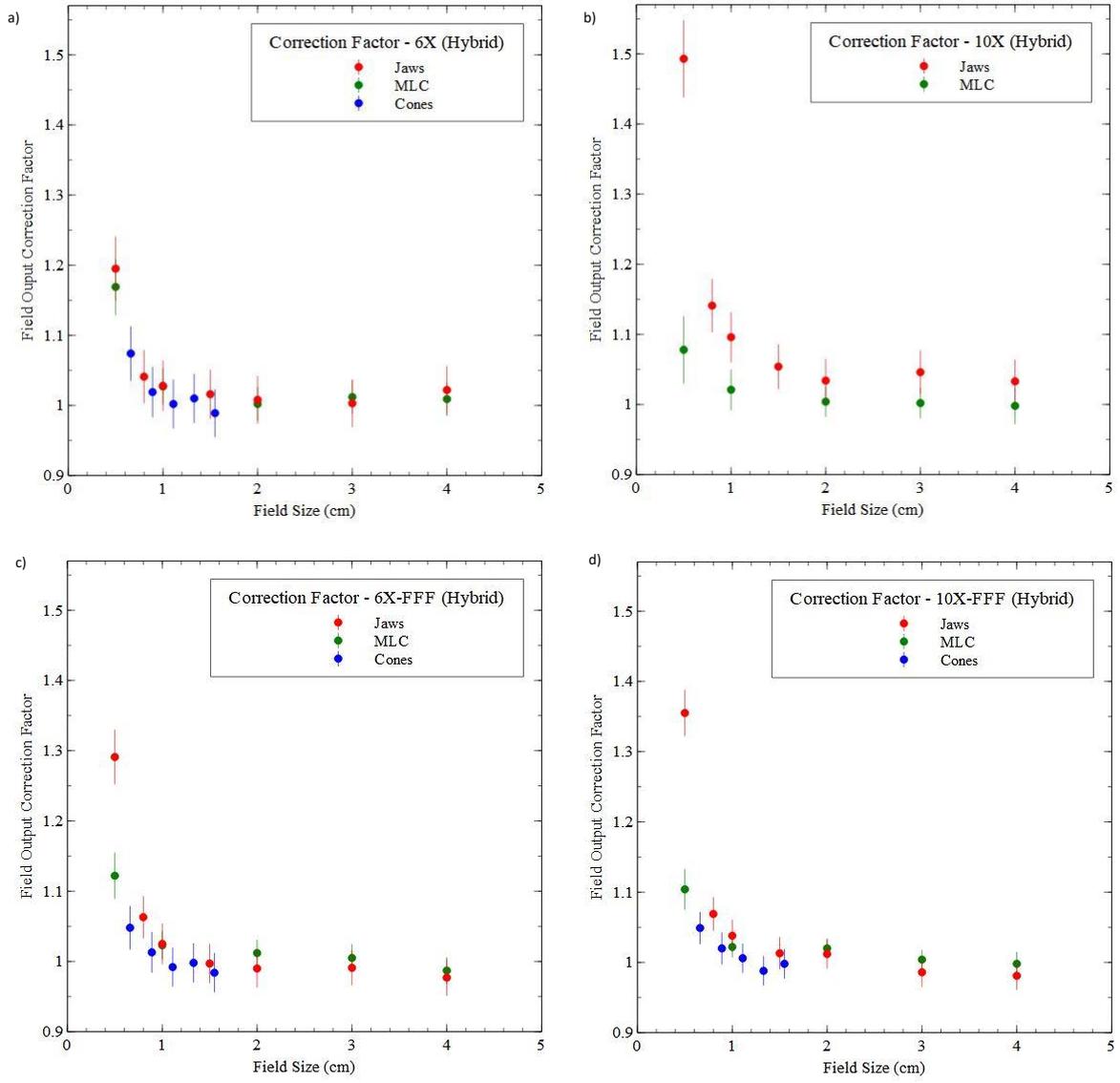

Figure 9 - $k_{Q_{clin},Q_{ref}}^{f_{clin},f_{ref}}$ for all the beam collimation acquired through the Hybrid Method: a) 6X, b) 10X, c) 6X-FFF and d) 10X-FFF.

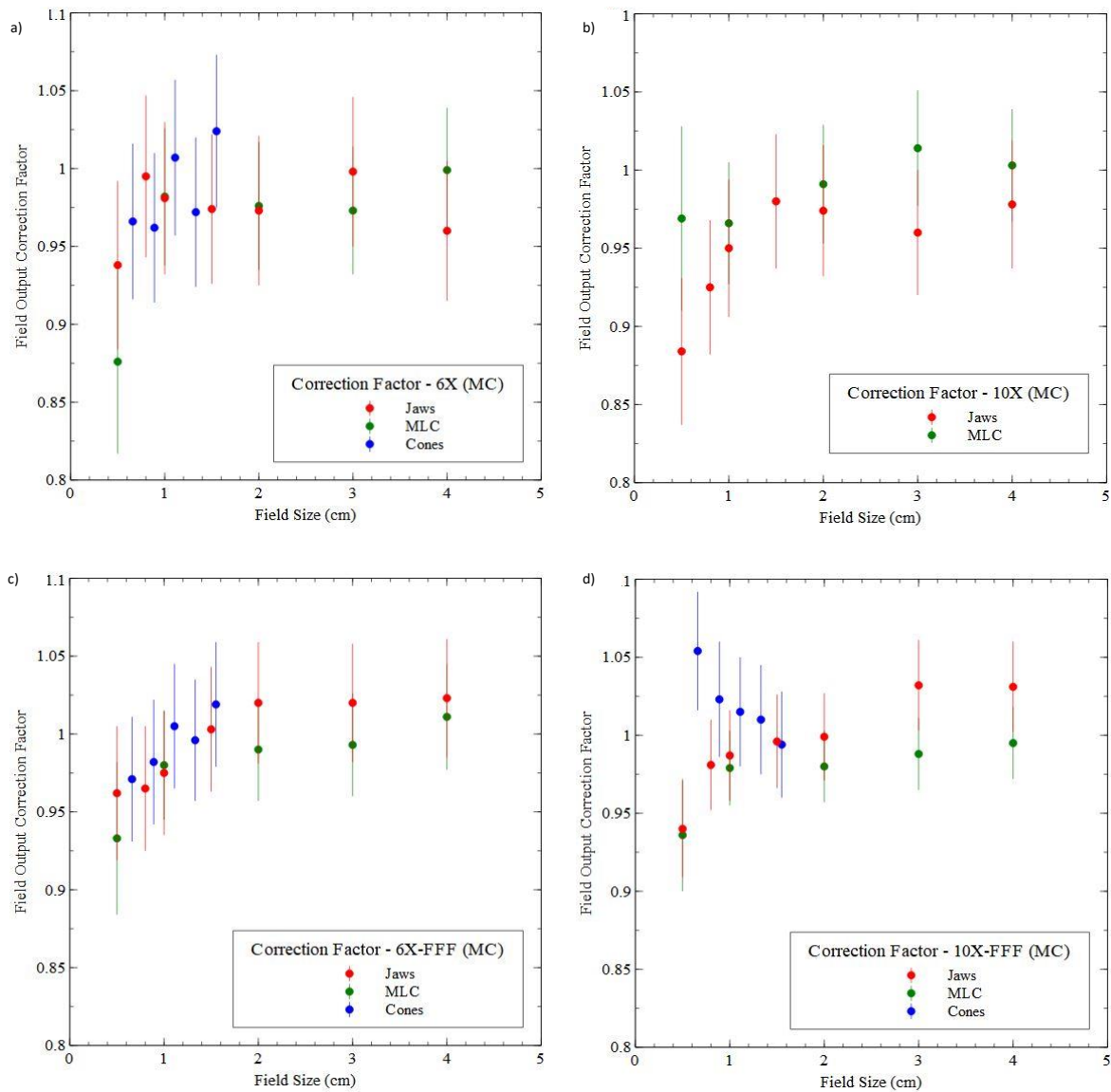

Figure 10 - $k_{Q_{clin},Q_{ref}}^{f_{clin},f_{ref}}$ for all the beam collimation acquired through the Monte Carlo Method: a) 6X, b) 10X, c) 6X-FFF and d) 10X-FFF.

## 4. Conclusions:

The IBA CC003 is a viable option for relative dosimetry for small field sizes (until 1x1 cm²). A significant improvement in spatial resolution compared to standard ionization chambers. For very small fields, however, the solid detectors (such as IBA Razor and PTW 60019) provide better results in terms of spatial resolution, which lead to reduced penumbra (around 1-2 mm less) and limited volume average effect. With respect to the field output factors, IBA CC003 can be used for small field sizes, but for field sizes $S_{clin} \leq 1$ cm it is necessary to apply field output correction factor to compensate for volume averaging and perturbation effects. The parallel orientation of the IBA CC003 seems to present more stable behavior when compared to the perpendicular orientation, because in perpendicular orientation the IBA CC003 can lead to stem effect issues, which were observed especially for field sizes below to 2x2 cm².

For the experimental determination of the field output factors using IBA CC003, it is important to take into account the polarity effect, especially for the parallel orientation of the ionization chamber, because the polarity effect increases with the decreasing of the field size. However, the ion collection efficiency is not as relevant, because the variation of ion collection efficiency with the field size is not significant. There is a difference between experimental and MC $k_{Q_{clin},Q_{ref}}^{f_{clin},f_{ref}}$, which increases with the decreasing of the field.

The major weakness of this simulation is the number of particles that reach the simulated ionization chamber, directly proportional to the field size, because the program is dependent on a limited particle file (phase-space file) provided by Varian.

The results for $k_{Q_{clin},Q_{ref}}^{f_{clin},f_{ref}}$ demonstrated the need of field output correction factors for equivalent square field sizes equal or less than 1 cm, for both orientations of the IBA CC003 with respect to the beam axis.

## 5. Acknowledgment:

We acknowledge Varian for the supply of the phase-space files of TrueBeam accelerator and IBA Dosimetry and Sociedade Avanço for putting at our disposal the Razor NanoChamber and Razor diode. One of the authors (Mateus D) acknowledges the financial support of Mercurius Health. We also thank Ashley Rose Peralta for the review of the English text.